\documentstyle[aps,prl,epsfig]{revtex}

\begin{document}

\draft

\widetext

\title{Separation of time-scales and reparametrization invariance for aging
systems}

\author{Claudio Chamon$^{a}$, Malcolm P. Kennett$^{b}$, Horacio E.
  Castillo$^{a}$, and Leticia F. Cugliandolo$^{c,d}$}

\address{ $^a$ Department of Physics, Boston University, Boston, MA 02215 \\
$^b$ Department of Physics, Princeton University, Princeton, NJ 08544 \\ $^c$
Laboratoire de Physique Th{\'e}orique de l'Ecole Normale Sup{\'e}rieure, 75231
Paris Cedex 05, France \\ $^d$ Laboratoire de Physique Th{\'e}orique et Hautes
Energies, Jussieu, 75252 Paris Cedex 05, France}


\twocolumn[\hsize\textwidth\columnwidth\hsize\csname@twocolumnfalse\endcsname
\maketitle

\begin{abstract}
  We show that the generating functional describing the slow dynamics of spin
glass systems is invariant under reparametrizations of the time.  This result
is general and applies for both infinite and short range models.  It follows
simply from the assumption that a separation between short time-scales and
long time-scales exists in the system, and the constraints of causality and
unitarity.  Global time reparametrization invariance suggests that  
the low action excitations in a spin glass may be smoothly spatially varying 
time reparametrizations.
These Goldstone modes may provide the basis for an analytic 
dynamical theory of short range spin glasses. 

\end{abstract}

]


\narrowtext
Much progress has been made in recent years in understanding the
non-equilibrium dynamics of glassy systems. Late after a quench to low
temperatures, two distinct dynamic regimes develop \cite{BCKM,CugKurA}. At
short time-dif\-fe\-ren\-ces the dynamics is similar to that in the
disordered phase. The correlations depend only on time differences [time
translation invariance ({\sc tti})] and decay towards a nonvanishing
Edwards-Anderson ({\sc ea}) order parameter.  
The correlations and their associated
responses are related by the fluctuation-dissipation theorem ({\sc fdt}).  At
long time-dif\-fe\-ren\-ces, for a freely relaxing system, physical
quantities relax very {\it slowly} and depend on the waiting time, so that
correlations depend on two times rather than on time differences.  The
separation of time-scales exists also when a glassy system is gently
driven, whether the full dynamics becomes stationary or not.  In the slow
regime, the {\sc fdt} is replaced by an out-of-equilibrium fluctuation
dissipation relation ({\sc oefdr}) between each bulk response $R$ and its
associated bulk correlation $C$:
\begin{equation}
R(t_1,t_2) = \beta_{\sc eff}[C(t_1,t_2)]\; \frac{\partial}{\partial
t_2}C(t_1,t_2) \; ,
\label{hola}
\label{eq:modified-FDT}
\end{equation}
where $t_1\geq t_2$, and $\beta_{\sc eff}[C]$ is the inverse effective
temperature measuring deviations from the {\sc fdt} ({\it e.g.}
\cite{CugKurC}).  These properties, observed in experiments and
simulations~\cite{BCKM}, are embodied in the concepts of weak ergodicity
breaking and weak long term memory \cite{CugKurA,Bouchaud,CugKurB}.  This
scenario has been demonstrated analytically for mean-field glassy models, 
such as the Sherrington-Kirkpatrick model \cite{CugKurB}. However,     
for short-range models such as the 
Edwards-Anderson model \cite{EA}, no analytic solution is known
for either the statics or the dynamics.
The work here presented uncovers a symmetry that strongly constraints 
the properties of such a dynamic solution.

The {\sc oefdr} is consistent with the observation that the equations of
motion of a large class of glassy models are invariant under time
reparametrizations, $t\to h(t)$ in the slow
regime~\cite{CugKurC,CugKurB,FM,KC1,old}.  In particular,
\begin{eqnarray}
  \label{eq:C-trans}
  \tilde C(t_1,t_2) &=& C(\,h(t_1),\, h(t_2)\,) \; ,\\
  \label{eq:R-trans}
  \tilde R(t_1,t_2) &=& \left(\frac{\partial h}{\partial t_2}\right)
  R(\,h(t_1),\, h(t_2)\,) \; ,
\end{eqnarray}
are related by the same {\sc oefdr} as in Eq.~(\ref{eq:modified-FDT}).
Equations~(\ref{eq:C-trans}) and (\ref{eq:R-trans}) are particular cases of
the general correlator Reparametrization Group (R$p$G) transformation
\begin{equation}
  \label{eq:G-trans}
  \tilde G(t_1,t_2) = \left(\frac{\partial h}{\partial
  t_1}\right)^{\Delta_{A}^G} \left(\frac{\partial h}{\partial
  t_2}\right)^{\Delta_{R}^G} G(\,h(t_1),\, h(t_2)\,) \; ,
\end{equation}
with $\Delta_A^G$ and $\Delta_R^G$ defined in Ref.\cite{KC1} as,
respectively, the advanced and retarded scaling dimensions of $G$ under the
reparametrization $t\to h(t)$ of $t_{1,2}$. In particular, $\Delta^C_{A,R}=0$,
$\Delta^R_A=0$, and $\Delta^R_R=1$ [see Eqs.~(\ref{eq:C-trans}) and
(\ref{eq:R-trans})].  

In this paper we show that: i) R$p$G invariance exists at the level of the
generating functional for the long-time dynamics, and ii) it holds for {\it
  short-range} models and their {\it local}, and not only bulk, quantities.
These are much stronger results than those of
Refs.~\cite{CugKurC,CugKurB,FM,KC1,old}, which only applied to mean field
models and were restricted to global quantities.  Furthermore, the existence
of a continuous symmetry in the action allows us to predict the presence of a
Goldstone mode controlling the dynamics, which may be the basis for
developing a systematic analytical theory of dynamical fluctuations in short
range spin glasses.

In order to prove our result, we take 
three steps: i) we determine the long-time action by using the
Renormalization Group (RG) 
in the {\em time} variables, assuming that there is a separation between
short and long time scales, 
ii) we analyze the surviving terms in the action,
showing that they are R$p$G invariant, and iii) we show that the measure in
the functional integral is also R$p$G invariant.

For the sake of concreteness we discuss the general disordered spin model
defined by
\begin{equation}
\label{eq:H}
H=\sum_{ij} J_{ij} \; S_i S_j + H_{\rm loc} \; ,
\end{equation}
where $H_{\rm loc}$ contains arbitrary self-interactions that, for
example, place soft or hard ($\pm S$) constraints on the spins.  Typically,
the random couplings $J_{ij}$ are Gaussian distributed according to
$P(J_{ij})\propto 
\exp[-J^2_{ij}/(2K_{ij})]
$, with $K_{ij}$ the connectivity
matrix. In the EA model, $K_{ij}=\tilde J^2/z$ if $i,j$ are nearest
neighbors or zero otherwise ($z$ is the coordination of the lattice).  We
consider a quantum extension~\cite{KC1,CugLoz}, with quantization rules
imposed via a kinetic term, $\sum_{i=1}^N p_i^2/(2M)$, where the momenta
$p_i$ are canonically conjugate to the ``coordinates'' $S_i$, $[p_i,
S_j]=-i\hbar \delta_{ij}$.  The use of bosonic variables to represent the
spins ({\it e.g.} quantum rotors, which lack Berry phases) simplifies the
presentation considerably although the arguments below can be extended to
other glassy systems such as the SU(N) Heisenberg model~\cite{Bipa}.  The
system is linearly coupled to an equilibrated environment.

We study the dynamics with the Schwinger-Keldysh 
closed time-path formalism ~\cite{KC1,CugLoz,Bipa,Weiss,LuYu,CLN}
in which the 
spin variables acquire another index labeling the two Keldysh branches,
$S_i\to S_i^a$, $a=0,1$. We work in the rotated basis
$\sigma_i^0=\hat\sigma_i=(S_i^0 - S_i^1 )/\sqrt{2}$ and 
$\sigma_i^1=\sigma_i=(S_i^0 + S_i^1 )/\sqrt{2}$.
This
notation renders the classical limit, $\hbar\to 0$, more
transparent~\cite{CugLoz}.

Once the disorder has been integrated out, we introduce a {\it local}
Hubbard-Stratonovich field $Q_i(t_1,t_2)$~\cite{Sozi}
\begin{equation}
Q_i = \left( \matrix{ Q_i^{00} & Q_i^{01} \cr Q_i^{10} & Q_i^{11} \cr }
\right ) = \left( \matrix{ Q_i^{K} & Q_i^{R} \cr Q_i^{A} & Q_i^{D} \cr }
\right ) ,
\label{Qmatrix}
\end{equation}
that decouples the four spin term generated by the average over the $J_{ij}$.  The choice of indices $0,1$ 
anticipates the result that we 
prove, namely 
invariance of
the action under the transformation $ Q^{ab}_i(t_1,t_2) \rightarrow \tilde
Q^{ab}_i(t_1,t_2)$, with 
\begin{equation}
\label{eq:Q-transf}
\tilde Q^{ab}_i(t_1,t_2) = \left(\frac{\partial h}{\partial t_1}\right)^a
  \left(\frac{\partial h}{\partial t_2}\right)^b Q^{ab}_i(\,h(t_1),\,
  h(t_2)\,) 
\end{equation}
[{\it cf.} Eq.~(\ref{eq:G-trans})].  This {\it Ansatz} for how the $Q$ fields
transform is motivated as follows. The $Q^K$ component is a time-dependent
local measure of freezing (analogous to the Edwards-Anderson order parameter
for the static case) since it is related to the correlator at the same site
but at two different times; hence the choice of vanishing dimensions
$\Delta^{Q^K}_{A,R}=0$ for this slowly decaying quantity. The scaling
dimensions for the retarded and advanced components $Q^{R,A}$ are suggested
by the {\sc fdt}, once the dimensions of $Q^K$ are fixed.

In the usual Keldysh approach, one keeps only the three non-vanishing
components, $K,R,A$, since Keldysh propagators are related to
expectation values $\langle Q_i^{K} \rangle$, $\langle Q_i^{R}
\rangle$ and $\langle Q_i^{A} \rangle$, while $\langle Q_i^{D}
\rangle=0$ \cite{Weiss,LuYu}. Here we include the fourth component
because we are also interested in the fluctuations of the $Q_i$'s.
The generating functional reads
\begin{eqnarray}
& & Z = \int [DQ]\; \exp\left(-S_K[Q] - S_{\sc nl}[Q]\right) \;,
\label{eq:Zlocal}
\\ && S_K[Q]= \frac{1}{2} \sum _{ij} K^{-1}_{ij}\int dt_1 dt_2 \sum_{ab}
\;Q^{ab}_{i}(t_1,t_2)\; Q^{\bar{b} \bar{a}}_{j}(t_2,t_1) , \nonumber \\ &&
S_{\sc nl}[Q] = - \ln \int [D\sigma] [D\hat\sigma]\,
\exp\left({iS[\sigma,\hat\sigma;Q] + iS_{\sc spin}}\right) ,
\label{ZQ}
\\ && S[\sigma,\hat\sigma;Q] = \sum_{i,a,b} \int dt_1 dt_2 \;
\sigma^a_{i}(t_1)\; Q^{ab}_{i}(t_1,t_2)\; \sigma^b_{i}(t_2) \; .
\label{SsigmaQ}
\end{eqnarray}
$S_{\sc spin}$ includes all the {\it local} spin dynamics, including the
kinetic term, the self-interactions in $H_{\rm loc}$, and the coupling to the
bath. The overline is a shorthand notation such that $\bar 0=1$ and $\bar
1=0$.  All integrals start at $t=0$.  In order to observe nonequilibrium
glassy dynamics, we first take the thermodynamic limit, $N\to\infty$,
consider times such that the upper limit in the time-integrals diverges
subsequently~\cite{CugKurC}, and use a weak coupling to the bath.

\begin{figure} 
\epsfxsize=2.in \center \epsfbox{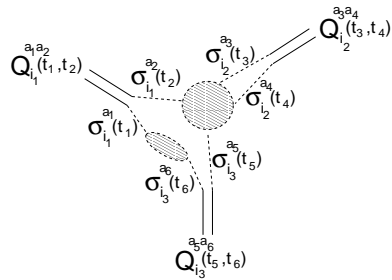} \vspace{.2cm}
\caption{A cubic term contributing to $S_{\sc nl}[Q]$. The
dashed lines are $\sigma$'s, which together with the shaded regions
represent connected Green's functions for the $\sigma$'s.  }
\label{fig:example1} 
\end{figure}

Upon integration over the spin variables $\sigma_i^a$ in Eq.~(\ref{ZQ}):
\begin{eqnarray}
& & S_{\sc nl}[Q] \! = \! \sum_{n=1}^\infty \frac{1}{n!}  
\sum_{i_1,\dots,i_n} \! \int \! dt_1 \dots dt_{2n} 
\;G^{i_1,\dots,i_n}_{a_1,\dots,a_{2n}}(t_1,\dots,t_{2n})
\nonumber \\ && \;\;\;\;\;\;\;\;\;\;\;\; \times \;Q^{a_1a_2}_{i_1}(t_1,t_2)
\cdots \;Q^{a_{2n-1}a_{2n}}_{i_n}(t_{2n-1},t_{2n}) \; ,
\label{eq:expansion}
\end{eqnarray}
with $G^{{i_1,\dots,i_n}}_{a_1,\dots,a_{2n}}(t_1,\dots,t_{2n}) \equiv
\langle \sigma_{i_1}^{a_1}(t_1) \cdots \sigma_{i_n}^{a_{2n}}(t_{2n})
\rangle$.  [$\langle \; \cdot \; \rangle$ denotes an average over the
fields $\sigma, \hat\sigma$ with the weight $\exp(i S_{\sc
spin})$].  The non-linear terms in Eq.~(\ref{eq:expansion}) are generated
because $H_{\rm loc}$ is non-quadratic. 
This expansion contains only
connected terms in the sense that none of the powers of $Q$ can be
factored. 
Each $Q$ serves as a source for a pair of $\sigma$'s, hence the $2n$-point
correlations $G^{(2n)}$ are 
not necessarily connected in terms of {\em individual $\sigma$'s}.  
In general, the contribution at any order
$n$ in $Q$ can be 
evaluated by rewriting the $G^{(2n)}$ into sums over
products $\prod_{\alpha=1}^M\;G_c^{(2m_\alpha)}$ of connected Green's
functions $G_c^{(2m_\alpha)}$, with $m_1+\cdots+m_M=n$. An example,
to order $Q^3$, is shown in Fig.~\ref{fig:example1}; notice that the
three $Q$'s form a connected bubble diagram, though at the same time
it contains both a 2-point and a 4-point connected Green's function
for the $\sigma$'s. Later on we will use this example to illustrate 
some of the steps in our arguments. 

i)  {\it RG in the time variable} - We assume there is an initial
short-time cutoff $\tau_0 = \Omega^{-1}$. We then separate the fast
and slow components of the field $Q_i$,
\begin{eqnarray*}
{Q_i^{ab}} (t_1,t_2) &=&
\cases{
{Q_i^{ab}}_{\sc fast} (t_1,t_2) \;\;\;{\rm if}\;\; 
\tau_0<|t_1-t_2|\le b\tau_0 \; , 
\cr
{Q_i^{ab}}_{\sc slow} (t_1,t_2) \;\;\,{\rm if}\;\; 
b\tau_0<|t_1-t_2|
\; ,
\cr
}
\end{eqnarray*}
for a rescaling $b>1$ (and $\delta \ell=\ln b>0$).  The couplings
that flow in the RG are the coefficients
$G^{i_1,\dots,i_n}_{a_1,\dots,a_{2n}}(t_1,\dots,t_{2n})$ in
Eq.~(\ref{eq:expansion}). As usual, each RG transformation involves two
operations. The first 
is the integration over the fast modes ${Q_i}_{\sc fast}$, which we perform 
by representing $S_{\sc nl}$ 
as a path integral over spin variables $\sigma_i^a$ 
[Eqs.~(\ref{ZQ}) and (\ref{SsigmaQ})].  The second 
is 
rescaling 
time $t\to
t/b$ 
to restore the cut-off to its original scale
$\tau_0=\Omega^{-1}$, accompanied by a rescaling of the $Q$ fields, 
performed on the expression in Eq.~(\ref{eq:expansion}). Each of these
operations generates a change of order $\delta \ell$ in the action.  
The quadratic term in the action, $S_K[Q]$, plays a role analogous to the
kinetic energy term in a usual RG calculation: it does not mix fast and slow
modes, and the rescaling of fields is chosen so as to keep it invariant.

The integration over fast modes yields a non-local four spin interaction
similar to the one obtained by performing the disorder average, but
fundamentally different in that it is {\it short} ranged in time. 
This extra spin-spin interaction leads to a 
change in the $2n$-point spin correlation functions, 
$\delta G^{(2n)}=\delta \ell \; [{d G^{(2n)}}\!/{d \ell}]|_{\rm fast}$.

As mentioned above, each coefficient $G^{(2n)}$ can be expressed as a sum of
products of connected Green's functions. 
Consider one of these
connected Green's functions, which plays the role of a coupling
for the $Q$'s.
In the same way that in a usual RG calculation (in real space) nonlocal
interactions flow into local interactions (i.e. the only couplings that are
left are those connecting fields at the same point), 
in our case all
couplings connecting different times flow into 
couplings that only connect {\em equal} times (i.e. other terms are
irrelevant compared to these terms).
Using this property, the term in Fig.~\ref{fig:example1} reads:
\begin{eqnarray}
&& \int \!d\tau
\; {G_c}^{i_1,i_3}_{a_1,a_6}(\tau,0)
\int \!d^{3}\tau
\; {G_c}^{i_1,i_2,i_2,i_3}_{a_2,a_3,a_4,a_5}(\tau',\tau'',\tau''',0)
\nonumber\\ 
&& \times 
\int dt_1 dt_2 \;Q^{a_1a_2}_{i_1}(t_1,t_2) \;Q^{a_3a_4}_{i_2}(t_2,t_2)
\;Q^{a_5a_6}_{i_3}(t_2,t_1) 
\; . \nonumber
\end{eqnarray}

Here and in the next equation, the $\tau$ variables represent time
differences (time can be shifted because of {\sc tti}) and their range of
integration is extended to be $(-\infty,+\infty)$.  
For a general term in the slow action, we obtain
an integral of a product of $Q$ fields
over time variables (one time variable per connected spin Green's function),
with prefactors given by the integrals of the connected spin Green's function
over all possible time differences that
define the coupling
constants:
\begin{eqnarray}
\label{eq:g}
{g}^{i_1,\dots,i_{2m}}_{a_1,\dots,a_{2m}} &&\equiv
\int \!d^{2m-1}\tau 
\;{G_c}^{i_1,\dots,i_{2m}}_{a_1,\dots,a_{2m}}(\tau_1,\dots,\tau_{2m-1},0)
\nonumber \\
&&=\lim_{\{\omega\}\to 0} \; {\chi_c}^{i_1,\dots,i_m}_{a_1,\dots,a_{2m}} 
(\omega_1,\dots,\omega_{2m-1})
\; .
\end{eqnarray}
They
correspond to physical zero-frequency ({\sc dc}) generalized
correlators ${\chi^{\sc dc}_c}$ of the spin variables.

Let us now turn to the effect of restoring the cut-off to the
original scale by rescaling the times, $t \to t'=t/b=t e^{-\delta\ell}$, and
the fields, $Q_i^{a_1a_2}(t_1,t_2)\to {Q'}_i^{a_1a_2}(t'_1,t'_2)=
b^{a_1+a_2}\;Q_i^{a_1a_2}(t_1,t_2)$. It is easy to see that it
satisfies two conditions: it produces the same effect as an R$p$G
transformation with $h(t)=bt$ [{\em cf.} Eq.~(\ref{eq:Q-transf})], and it
leaves the quadratic term $S_K$ invariant. In the example:
\begin{eqnarray}
 I_{\rm ex}[Q] & \equiv &\int dt_1 dt_2 \;{Q}^{a_1a_2}_{i_1}(t_1,t_2)
\;{Q}^{a_3a_4}_{i_2}(t_2,t_2) \;{Q}^{a_5a_6}_{i_3}(t_2,t_1)
\nonumber\\ 
 \;\; & = & b^2 \int dt'_1 dt'_2 \;\;b^{-(a_1+a_6)}\;b^{-(a_2+a_3+a_4+a_5)}\;
\nonumber\\ && 
\;\;\;\;\; \times 
\;{Q'}^{a_1a_2}_{i_1}(t'_1,t'_2)
\;{Q'}^{a_3a_4}_{i_2}(t'_2,t'_2) \;{Q'}^{a_5a_6}_{i_3}(t'_2,t'_1)
\nonumber\\ 
\;\; & = & b^{1-(a_1+a_6)} \; b^{1-(a_2+a_3+a_4+a_5)}\; I_{\rm ex}[Q'] \; .  
\nonumber
\end{eqnarray}
In a general term, for each time variable $t_\alpha$
there is 
a coupling
$g_\alpha \equiv
g^{i_1,\dots,i_{2m_\alpha}}_{a_1,\dots,a_{2m_\alpha}}$, and also  
an exponent $\Delta_\alpha \equiv a_1+\cdots+a_{2m_\alpha} \ge 0$ 
originating from the rescaling of fields. In the example $\Delta_1=
a_1+a_6$ and $\Delta_2=a_2+a_3+a_4+a_5$.

Combining the contributions from both the integration of fast modes and the
rescaling, we obtain the following flow equation for the coupling constant
$g_\alpha$:
\begin{equation}
\label{eq:RG}
\frac{d g_\alpha}{d\ell} =
(1-\Delta_\alpha)g_\alpha + \frac{d g_\alpha}{d\ell}\Big|_{\rm fast}
\end{equation}
Terms with $\Delta_{\alpha} = 0$ occur only when all the $a$'s are 0. These
terms would be naively relevant (according to their engineering
dimensions), but in fact they vanish identically due to the constraints of
normalization and causality for $G_c^{(2n)}$ \cite{Weiss,LuYu}:
\begin{eqnarray*}
G^{i_1,\cdots,i_{2n_\alpha}}_{0,0,\dots,0}(t_1,\dots,t_{2n_\alpha})= \langle
\hat\sigma_{i_1}(t_1) \cdots \hat\sigma_{i_{2n_\alpha}}(t_{2n_\alpha})
\rangle =0 \; .
\end{eqnarray*}

There remain only terms which are either marginal (for $\Delta_{\alpha}
= 1$) or irrelevant (for $\Delta_{\alpha} \ge 2$) according to their
engineering dimensions. Under the assumption of a 
separation between short and long time
scales, at some point in the RG flow all of the
fluctuations associated with short time scales will have been integrated
over, (but the flow will not have yet reached a point were the long time
scales are probed) and therefore any new integrations of fast modes produce
no change in the coupling constants, i.e. the second term in the r.h.s. of
Eq.~(\ref{eq:RG}) is zero. This assumption is used in an analogous way in the
solution of mean field spin glass models~\cite{BCKM,CugKurA,CugKurC}.
Physically, the separation between short and long time scales correspond to a
saturation of the generalized {\sc dc} correlators
for the 
spins ${\chi^{\sc dc}_c}$ defined in Eq.~(\ref{eq:g}) to a {\it finite} value. 
Therefore, the existence of a separation
of time scales implies that the engineering dimension actually determines the
long time behavior, and only marginal ($\Delta_{\alpha} = 1$) terms are left
in the effective long-time action.

ii) {\it R$p$G invariance of the effective long-time action.} 
For a R$p$G transformation applied to a generic term in $S_{\sc nl}$, each
integral over time gives rise to a factor 
$\left({\partial h(t_{\alpha})}/{\partial t_{\alpha}}\right)^{\Delta_\alpha}$,
and since only terms with $\Delta_{\alpha}=1$ are present, the derivative
factor yields exactly the Jacobian necessary to make the term R$p$G invariant:
\begin{eqnarray*}
\int \prod_{\alpha} dt_\alpha \cdots & \to & 
\int \prod_{\alpha} dt_\alpha 
\left(\frac{\partial h}{\partial t_\alpha}\right)^{\Delta_\alpha} \cdots \\
& = & \int \prod_{\alpha} dt_\alpha 
\left(\frac{\partial h}{\partial t_\alpha}\right) \cdots 
= \int  \prod_{\alpha} dh_\alpha \cdots \; .
\end{eqnarray*}
To illustrate this point, let us perform a R$p$G
transformation on our example term:
\begin{eqnarray}
&& I_{\rm ex}[\tilde Q] = \int dt_1 dt_2 
\;{\tilde Q}^{a_1a_2}_{i_1}(t_1,t_2)
\;{\tilde Q}^{a_3a_4}_{i_2}(t_2,t_2) \;{\tilde Q}^{a_5a_6}_{i_3}(t_2,t_1)
\nonumber\\ 
&& =\int dt_1 dt_2 
\left({\partial h(t_1)}/{\partial t_1}\right)^{a_1+a_6} 
\left({\partial h(t_2)}/{\partial t_2}\right)^{a_2+a_3+a_4+a_5} 
\nonumber\\ 
&& \times 
\;Q^{a_1a_2}_{i_1}(h(t_1),h(t_2)) \;Q^{a_3a_4}_{i_2}(h(t_2),h(t_2))
\;Q^{a_5a_6}_{i_3}(h(t_2),h(t_1)) 
\nonumber\\ 
&& =\int dh_1 dh_2 
\;Q^{a_1a_2}_{i_1}(h_1,h_2) \;Q^{a_3a_4}_{i_2}(h_2,h_2)
\;Q^{a_5a_6}_{i_3}(h_2,h_1) \nonumber\\
&&= I_{\rm ex}[Q] 
\; ; \nonumber
\end{eqnarray}
where we have used that $a_1+a_6=\Delta_1=1$ and
$a_2+a_3+a_4+a_5=\Delta_2=1$. 
For the $S_K$ term, the same argument applies with $\Delta=1$ replaced by
$a+\bar{a}=1$ and $b+\bar{b}=1$. Therefore the long-time action is R$p$G
invariant. 

iii) {\it R$p$G invariance of the measure} - Under the R$p$G
transformation of Eq.~(\ref{eq:Q-transf}), the Jacobian for the
functional integral over the $Q$ fields is simply:
\begin{eqnarray*}
{\cal J}\left[\frac{D{\tilde Q}}{DQ}\right] = 
\prod_x \prod_{t_1,t_2}\;\left|\frac{\partial h}{\partial t_2} 
                \frac{\partial h}{\partial t_1}\right|^2 
= 
e^{\int d^dx dt_1 dt_2 \; \ln \left|\frac{\partial h}{\partial t_1} 
                \frac{\partial h}{\partial t_2}\right|^2}
\; .
\end{eqnarray*}
The Jacobian depends on $h(t)$, but not on
the fields $Q$, and therefore the generating functional is R$p$G
invariant. 

The R$p$G invariance
implies that the action describing the long time slow dynamics of a
spin-glass is basically a ``geometric'' random surface theory
\cite{Polyakov}, with the $Q$'s themselves 
as the natural coordinates. The
original two times parametrize the surface. Physical quantities, as the 
bulk integrated response $\chi(t_1,t_2)=\int_{t_2}^{t_1} dt' \, \langle
Q^R(t',t_2) \rangle $ and correlation $\langle Q^K\rangle(t_1,t_2)$
have scaling dimension zero under $t\to h(t)$ \cite{CugKurB,KC1} as
well as their local counterparts that are 
directly related to the $Q_i$'s. A
possibly related gauge-like symmetry has also been noted in the replica
approach \cite{replica-inv}.

The emergence of the R$p$G invariance may provide a
novel, 
completely
dynamical, angle to address the still poorly understood short-range
spin-glass problem analytically. We have shown that the {\it global}
reparametrization, $t\to h(t)$, is a symmetry of the slow dynamical
action. The particular scaling function $h(t)$ selected by the system is
determined by matching the fast and the slow dynamics. It depends on several
details -- the existence of external forcing, the nature of the microscopic
interactions, etc. In other words, the fast modes which are absent in the
slow dynamics act as symmetry breaking fields for the slow modes.  The {\it
global} 
R$p$G invariance of the slow action suggests that the low energy physics of
the glassy phase could be described by slowly spatially varying
reparametrizations $t\to h(\vec{x},t)$.  Basically, we propose that there are
Goldstone modes for the glassy action which can be written as slowly varying,
spatially inhomogeneous time reparametrizations. Comparing with the $O(N)$
non-linear sigma model \cite{Brezin}, {\it global} time-reparametrizations
are analogous to uniform spin rotations, while local $t\to h(\vec{x},t)$
reparametrizations describe the spin waves (fluctuations on the uniform 
solution). Numerical tests in the
3D EA model are consistent with this conjecture~\cite{prep}.

This work was supported by NSF grants DMR-98-76208, INT-01-28922, and the A.
P. Sloan Foundation (C.C.), by ACI France and CNRS-12931 (L.F.C.).  C.C.
thanks the LPTHE and M.P.K thanks Boston Univ. for their hospitality. L.F.C.
is staff associate at ICTP.




\begin{references}

  
\bibitem{BCKM} J. P. Bouchaud {\it et al}
in {\it Spin Glasses and Random Fields} A. P. Young ed.
  (World Scientific, Singapore, 1998).

\bibitem{CugKurA} L. F. Cugliandolo and J. Kurchan, Phys. Rev. Lett. {\bf
  71}, 173 (1993); Phil. Mag. B {\bf 71}, 501 (1995).

\bibitem{CugKurC} L. F. Cugliandolo and J. Kurchan, Physica A{\bf 263}, 242
(1999).


\bibitem{Bouchaud} J. P. Bouchaud, J. Phys. I (France) {\bf 2}, 1705 (1992).

\bibitem{CugKurB} L. F. Cugliandolo and J. Kurchan, J. Phys. A {\bf 27}, 5749
(1994).

\bibitem{EA} S. F. Edwards and P. W. Anderson, J. Phys. F {\bf 5}, 965 (1975).

\bibitem{FM} S. Franz and M. M{\'e}zard, 
Physica A {\bf 210}, 48 (1994).
   
\bibitem{KC1} M. P. Kennett and C. Chamon, Phys. Rev. Lett.  {\bf 86}, 1622
(2001); M. P. Kennett, C. Chamon and J. Ye, Phys. Rev. B {\bf 64}, 224408
 (2001).

\bibitem{old} H. Sompolinsky, Phys. Rev. Lett. {\bf 47}, 935 (1981);
V. S. Dotsenko, M. V. Feigel'man and L. B. Ioffe, {\it Spin glasses and related
problems}, Soviet Scientific Reviews {\bf 15}, (Harwood, 1990).



\bibitem{CugLoz} L. F. Cugliandolo and G. Lozano, Phys. Rev. Lett. { \bf 80},
4979 (1998); Phys. Rev. B {\bf 59}, 915 (1999).

\bibitem{Bipa} G. Biroli and O. Parcollet, Phys. Rev. B {\bf 65}, 094414 (2002).

\bibitem{Weiss} U. Weiss, {\it Quantum dissipative systems} (World
Scientific, 1999).

\bibitem{LuYu} K.-C. Chou {\it et al.}
Phys. Rep.  {\bf 118}, 1 (1985).

\bibitem{CLN} C. Chamon, A. W. W. Ludwig, and C. Nayak, Phys. Rev. B {\bf
60}, 2239 (1999); A. Kamenev and A. Andreev, {\it ibid.} {\bf 60}, 2218
(1999).

\bibitem{Sozi} H. Sompolinsky and A. Zippelius
   Phys. Rev. Lett. {\bf 50}, 1297 (1983). 
%


  
\bibitem{Polyakov} A. M. Polyakov, {\it Gauge Fields and Strings}, (Harwood
  Academic Publishers, Chur, Switzerland, 1987).

\bibitem{replica-inv} I. Kondor and C. De Dominicis, Europhys. Lett.  {\bf
2}, 617 (1986); T. Temesv{\'a}ri {\it et al}
Eur. Phys. J. B {\bf 18}, 493 (2000).

\bibitem{Brezin} 
E. Br{\'e}zin and J. Zinn-Justin, Phys. Rev. B{\bf 14}, 3110
  (1976). 
  
\bibitem{prep} H. E. Castillo {\it et al}. Phys. Rev. Lett. {\bf 88}, 237201 (2002).


\end{references}
\end{document}